# Hot electron lifetime exceeds 300 nanoseconds in quantum dots with high quantum efficiency

Beibei Tang[1,2,5]†, Bo Li[1,2,5]†, Yingying Sun[3], Jianshun Li[3], Yanheng Guo[3], Jiaojiao Song[3], Xiaohan Yan[1,2,5], Huimin Zhang[3], Xiaosuo Wang[1,2,5], Fei Chen[3], Lei Wang[3], Jiangfeng Du[1,2,4,5], Huaibin Shen[3], Fengjia Fan[1,2,5]*

[1] CAS Key Laboratory of Microscale Magnetic Resonance and School of Physical Sciences, University of Science and Technology of China, Hefei 230026, China.

[2] Anhui Province Key Laboratory of Scientific Instrument Development and Application, University of Science and Technology of China, Hefei 230026, China.

[3] Key Laboratory for Special Functional Materials of Ministry of Education, National & Local Joint Engineering Research Center for High-efficiency Display and Lighting Technology, Henan University, Kaifeng 475004, China.

[4] Institute of Quantum Sensing and School of Physics, Zhejiang University, Hangzhou 310027, China.

[5] Hefei National Laboratory, University of Science and Technology of China, Hefei, 230088, China.

*E-mails: ffj@ustc.edu.cn;

†These authors contributed equally to this work.

## Abstract

**Hot electrons are theoretically predicted to be long-lived in strongly confined quantum dots, which could play vital roles in quantum dot-based optoelectronics; however, existing photoexcitation transient spectroscopy investigations reveal that their lifetime is less than 1 ps in well-passivated quantum dots because of the ultrafast electron-hole Auger-assisted cooling. Therefore, they are generally**

**considered absent in quantum dot optoelectronic devices. Here, by using our newly developed electrically excited transient absorption spectroscopy, we surprisingly observed abundant hot electrons in both II-VI and III-VI compound quantum dot light-emitting diodes at elevated bias (>4 V), of which the lifetimes reach 59 to 371 ns, lengthened by more than 5 orders of magnitude compared with the photoexcited hot electrons. These results experimentally prove the presence of a strong phonon bottleneck effect, refreshing our understanding of the role of hot electrons in quantum dot optoelectronics.**

## Main text

The phonon bottleneck is an intriguing physics phenomenon that is theoretically predicted to exist in strong quantum-confined nanostructures[1,2], particularly in colloidal quantum dots (QDs)[3,4]. The hot charge carriers relaxation among discrete energy levels is expected to be greatly retarded because the energy gap is way larger than the phonon energy[5,6]. Long-lived hot electrons could revolutionize solar cell technology because they allow the breaking of the Shockley-Queisser limit[7-9]. Their existence could also shed some new light on the electron behavior in optoelectronics devices[10,11], such as light-emitting diodes, and field effect transistors because the carriers might populate not only the band edge states but also the higher energy states, which are neglected in our current understanding.

During the last three decades, there have been many hot carrier investigations in search of phonon bottleneck using ultrafast photoexcited transient absorption (PETA) and time-resolved photoluminescence[12-17]. The results reveal that hot carriers relax to the band edge within femtoseconds to picoseconds, even in strongly confined nanocrystals[3,14,16]. This is because there are other faster processes, such as electron-hole Auger and trap state assisted cooling[18-20], happening before the anticipated slow multi-phonon emission. By introducing hole-trapping ligands and heavy n-doping to eliminate holes after photoexcitation, the hot electron lifetime can be extended to several hundred picoseconds to 1 nanosecond[5,21-23]. However, these QDs are full of trap states, and not suitable for optoelectronic applications. Therefore, hot electrons are generally considered absent in QD optoelectronic devices.

Here, we apply our latest electrically excited transient absorption (EETA) spectroscopy[24,25] to probe the lifetime of hot electrons. Unlike photoexcitation, which always generates electrons and holes in pairs, electrical excitation allows us to inject pure hot electrons in QDs without introducing holes[26]. Surprisingly, we found abundant long-lived hot electrons with lifetimes up to 59 and 371 ns in high quantum yield II-VI and III-V QDs, respectively, far exceeding the lifetime measured using ultrafast PETA spectroscopy (< 1 ps), demonstrating the long-anticipated phonon bottleneck effect. These results refresh our understanding of the potential and role of hot electrons in QD optoelectronic devices.

**The relaxation process of photoexcited excitons and electrically excited hot electrons**

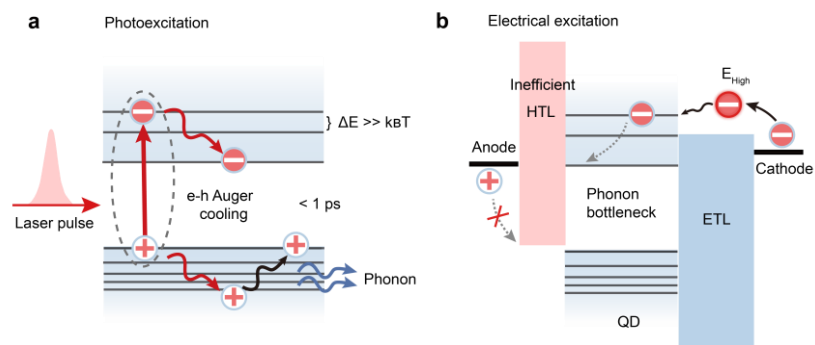

**Fig. 1 | Hot electron relaxation process. a**, Electron-hole Auger-assisted hot electron cooling process in photoexcited QDs. Hot electrons relax to the band edge by transferring excess energy to holes, which cool to the band edge by emitting phonons. **b,** The relaxation process of the hot electrons injected at high driving voltages. In the case of inefficient hole injection, the hot electron relaxation process is much slower.

In III-V and II-VI semiconductors, the density of states in the conduction band is low and the effective mass is small[27-29]. With strong quantum confinement, the electron energy levels are so spaced out that their interband transition through phonon emission is rather inefficient, which will, in principle, lead to the so-called phonon bottleneck, and long-lived hot electrons[4]. However, this theoretically predicted phenomenon has been hardly observed experimentally.

In previous ultrafast PETA spectroscopy and transient photoluminescence studies, the hot excitons in QDs relax to the band edge in picoseconds[3], showing almost no difference from the bulk counterpart. The explanation for this phenomenon is that there is no phonon bottleneck for the holes because the energy levels are still closely packed in the valence band[16], and the hot electrons can lose their energy through Coulomb coupling with holes (Fig. 1a), i.e. electron-hole Auger cooling. This explanation is supported by experiments in which holes are annihilated by the trapping ligands and introduced heavy n-doping, the hot electron lifetime can be lengthened by 3 orders of magnitude, reaching 1 nanosecond[5,23].

However, electron-hole Auger cooling is an extremely fast process, its effect cannot be completely eliminated in photoexcited transient spectroscopy, and therefore how long-lived could the pure hot electron be remains an unsolved puzzle. This is a question of not only fundamental scientific significance but also practical application interest. As in some optoelectronic devices, such as LEDs[30,31], and field effect transistors[32], hot electrons can be generated at higher driving voltages, but the holes could be either absent or occupying different QDs with electrons, i.e. there are QDs occupied by electrons only. This is also critical to evaluate the prospect of the hot carrier QD solar cells – if the electron-hole separation is fast enough[33], is it practically possible to extract hot electrons? However, due to the lack of proper instruments, this is no clear answer to these questions.

In this work, we unveil the lifetime limit of hot electrons in QDs using a newly developed EETA spectroscopy. We applied electrical pulses to pump high-energy electrons into the QD-LED, and then probed the change in absorbance using white light laser pulses with different time delays. In doing so, we can observe electron occupation and relaxation processes among different energy levels. By choosing hole transport layers (HTL) with low mobility and deep HOMO energy level[34], we can block hole injection, therefore, the relaxation path through electron-hole Auger cooling can be efficiently depressed, allowing us to evaluate the hot electron lifetime limit.

**The observation of hot electrons in CdSe-based QDs under electro-excitation**

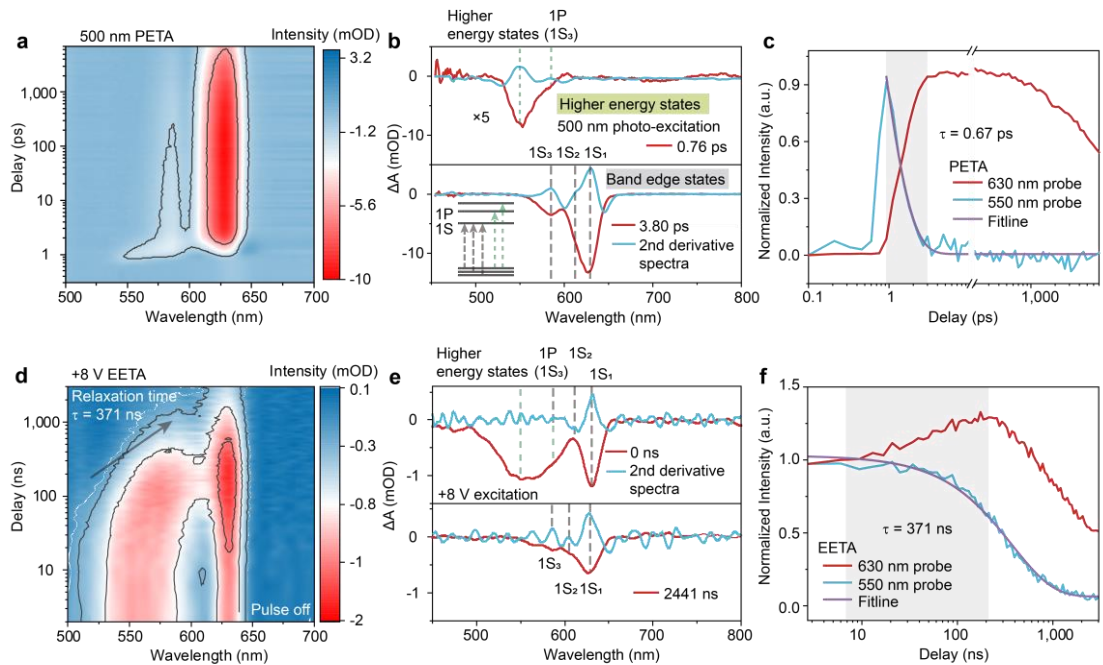

**Fig. 2 | Electron bleaching signals of CdSe-based QDs under photo- and electro-excitation. a,** PETA spectra of QDs dispersed in octane using 500 nm 190 fs lasers. **b,** PETA spectra (solid red line) with delay times of 0.76 ps (top panel) and 3.8 ps (bottom panel), respectively. The blue solid line is the second-order differential spectra, and the green and gray dashed lines represent the signals attributing to higher energy and band edge electron states, respectively. The inset shows the energy level diagram of the ZnCdSe/ZnSeS QDs. **c,** Bleaching signal kinetics at 550 nm and 630 nm extracted from Fig. 2a. **d,** EETA spectra with +8 V 10 μs electrical pulse collected in QD-LEDs (ITO/PEDOT:PSS/PVK/ZnCdSe/ZnSeS QDs/$Zn_{1-x}Mg_xO$/Al) with inefficient hole injection. **e,** EETA spectra collected right after the end of the electrical pulse (0 ns, top

panel), and after a delay time of 2441 ns (bottom panel), respectively. **f,** Kinetics of bleaching signal at 630 nm (solid red line) and 550 nm (solid blue line), respectively.

In our experiments, we first investigated ZnCdSe/ZnSeS QDs which show high photoluminescence and electroluminescence efficiency (Supplementary Fig. 1, 2 and Table 1), their surface is well passivated so the coupling with trap states is minimized. The electrons in higher energy states are still confined because of type-I band alignment.

We first performed PETA experiments (Fig. 2a-c and Supplementary Fig. 3) on QDs dispersion in octane to measure the hot exciton relaxation time and, more importantly, to resolve the band edge and higher energy states. These data will serve as a reference for analyzing the EETA spectra. Since the degeneracy of the hole energy level is much higher than that of the electron, the exciton bleaching signal is mainly contributed by the electrons[35]. 0.76 ps after photoexcitation (190 fs, 500 nm), we observed an asymmetric bleaching peak at 550 nm, primarily attributed to electrons occupying 1P and higher energy states of the QDs. After a delay time of 3.8 ps, the bleaching signal shifts to 630 nm, and the relative intensity of different peaks stabilizes, indicating the hot electrons have relaxed to the band edge. The second-order differential of these spectra reveals three peaks, which, according to the spectra obtained under near-resonance photoexcitation at 620 nm (Supplementary Fig. 4), are attributed to transitions between band edge 1S electrons and different hole states. By performing single exponential fitting on the decay dynamics at 550 nm, we obtained a hot electron lifetime of 0.67 ps, which matches the rising time of the band edge electron bleaching signal, revealing electrons transfer from the higher energy to band edge states.

We then analyzed the EETA spectra (Fig. 2d-f, and Supplementary Figs. 5, 6). Since the charge injection through electrical pumping is much slower than that of the photoexcitation, we applied long electrical pulses of 10 μs with the bias of 8 V to let the charges in QD-LED reach equilibrium and investigate the decay process after the pulse ended. We first analyzed the equilibrated bleaching signal (0 ns delay time). Besides the 630 nm peak, we observed a broad peak located at 550 nm attributed to the electrons occupying higher energy states, indicating the hot electron lifetime is of the same order of magnitude as that of the band edge electrons.

We then investigated the dynamics after the pulse is off, while the higher energy state signal keeps decaying, the band edge state signal rises in the first 200 ns and then starts to decay, representing the cooling process, as observed in PETA spectra (Fig. 2f). Fitting on the decay dynamics of the higher energy state electron signal at 550 nm yields a lifetime of 371 ns, which is more than five orders of magnitude longer than that observed under photoexcitation. The fitted band edge electron lifetime reaches 1 μs, indicating low trap densities of QDs.

All above signals originate from QDs, as they are absent without QDs, and they are not generated from the gap states between the HTL and QDs, because they are still there after inserting PMMA layers between them (Supplementary Fig. 7). The bleaching signal around 550 nm does not originate from the Stark effect, too, because the Stark signal is in a differential-like shape (Supplementary Figs. 8, 9), which is not the one in Fig. 2b. In addition, the decay lifetime of the Stark effect is around 66 ns (Supplementary Fig. 9c), which is about 5 times faster than that of a hot electron.

**Evidence of the phonon bottleneck effect in CdSe-based QDs**

The above results indicate there are long-lived electrons occupying high energy states, however, it could be due to two different reasons: the state filling and phonon bottleneck effect. In our experiments, we found two sets of evidence that point to the phonon bottleneck effect. The first set of evidence is from the rising edge of EETA spectra, the higher energy and band edge state bleaching signals rise simultaneously rather than appear sequentially (Supplementary Fig. 10). The second set of evidence is from the comparison between band edge bleaching of EETA and PETA spectra collected at the exact same spots (Fig. 3b, 3c and Supplementary Fig. 11, methods for more details). In EETA spectra, we observe an intense higher energy state signal when the band edge one reaches -1.7 mOD (+8 V); however, to get the same amplitude of the signal for the band edge state in PETA spectra, we only need to pump the QD with 0.47 excitons in average, well below the required population of 2 for band edge state filling (Supplementary Fig. 11).

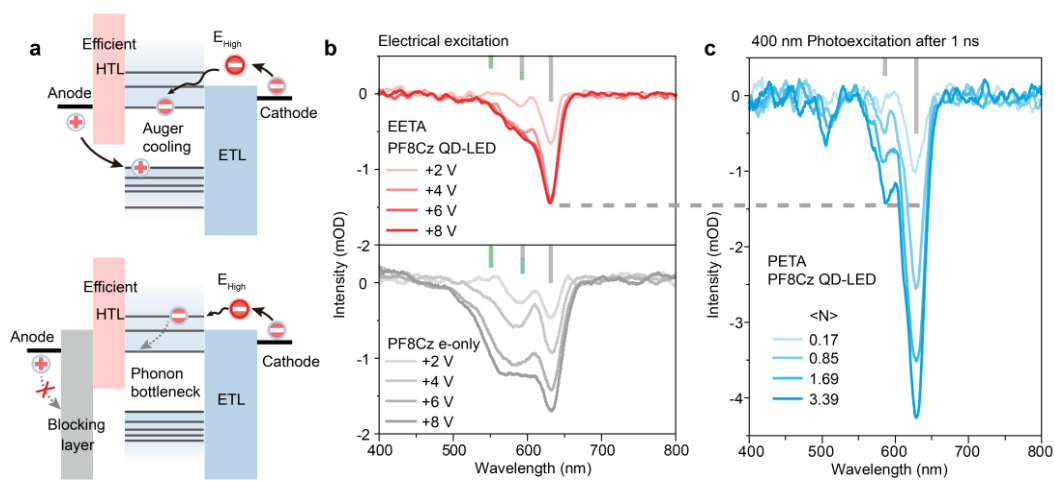

**Fig. 3 | Carrier population in operating CdSe-based QD-LEDs with different hole injection rates. a,** Upper panel: Schematic diagram of carrier injection in PF8Cz QD-LED (ITO/PEDOT:PSS/PF8Cz/ZnCdSe/ZnSeS QDs/Zn$_{1-x}$Mg$_x$O/Al) with high efficiency hole injection at high driving voltage. Lower panel: Schematic diagram of carrier injection in electron conductance-only QD-LED (ITO/ZnO/PF8Cz/ZnCdSe/ZnSeS QDs/Zn$_{1-x}$Mg$_x$O/Al). **b,** Steady EETA spectra collected for PF8Cz QD-LED (upper panel) and electron conductance-only QD-LED (lower panel), respectively. The solid green and gray lines represent the higher energy and the band edge states, respectively. **c,** PETA spectra obtained from PF8Cz QD-LED at the same spot where we collected EETA spectra, under photoexcitation of 400 nm laser with different pump fluences, the delay time is 1 ns.

The above experiment results suggest that by eliminating the electron-hole Auger cooling, the hot electron lifetime can reach an unbelievably long value. To further verify this conclusion, we adjusted the hole injection rate to see its impact on the hot electron population (Fig. 3). We first performed EETA measurements on QD-LEDs with more efficient HTL PF8Cz (Fig. 3a and 3b, upper panel), which featured higher hole mobility and a shallow HOMO level than those of PVK[34]. We found the hot electron population indeed dramatically decreases in these devices, and their weak presence is only observable at voltages above 4 V. We then measured devices still using PF8Cz but with hole injection blocked by ZnO (Fig. 3a and 3b, lower panel) (i.e. electron conductance only devices). We saw the reappearance of an intense hot electron signal. These consistent findings prove that the electron-hole interaction is extremely efficient in cooling hot electrons.

**The observation of hot electrons in InP-based QD-LEDs**

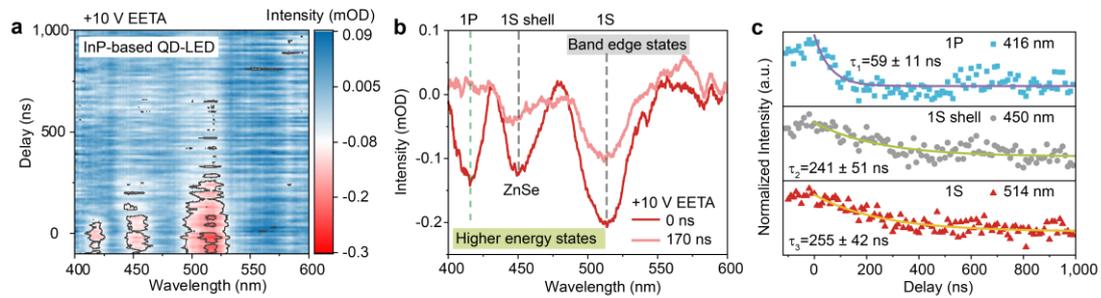

**Fig. 4 | The hot electron relaxation of InP-based QDs. a.** EETA spectra (electrical pulse: +10 V 10 μs) measured on InP-based QD-LEDs with inefficient hole injection (ITO/PEDOT:PSS/PVK/InP/ZnSe/ZnS QDs/$Zn_{1-x}Mg_xO$/Al). **b.** EETA spectra collected right after the end of the electrical pulse (0 ns, red line), and after a delay time of 170 ns (pink line), respectively. **c,** Kinetics of the bleaching signal peaking at 416 nm, 450 nm, and 514 nm, respectively.

To investigate whether the long-lived hot electron is specific to CdSe-based QDs or instead, widely exists in other QDs, we performed EETA analysis on InP/ZnSe/ZnS QD-LEDs using the same ETL and HTL. The steady-state EETA spectra (under electroexcitation of 10 V and 10 μs) show three sets of the bleach signals (416, 450 and 514 nm, respectively, Fig. 4a). To correlate these bleach signals to the electron states, we also performed PETA analysis (Supplementary Fig. 12). The highest energy one (416 nm) is attributed to hot electrons, as they disappear 1 ps after photoexcitation, while the two with lower energies (450 and 514 nm) are attributed to the band edge electrons because their lifetimes are both much longer (24 ± 1 and 27 ± 1 ns, respectively) under photoexcitation. Under electro-excitaion, the lifetime of band edge electrons reaches 255 ± 42 ns (Fig. 4c). More importantly, the hot electron relaxation process is also drastically retarded, reaching 59 ± 11 ns. These data demonstrate a strong phonon-bottleneck effect in InP-based QDs.

**The impact of hot electrons in QD-LEDs**

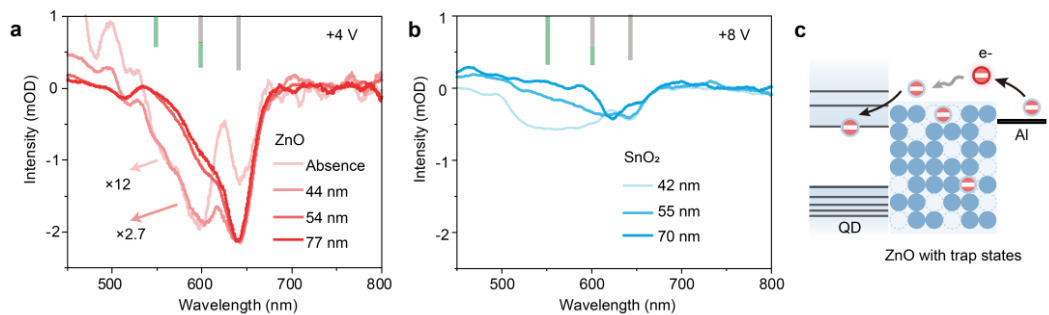

**Fig. 5| Carrier population in CdSe-based QD-LED using ZnO and SnO₂ nanoparticles as electron transport layers (ETLs).** EETA spectra collected on CdSe-based QD-LEDs (ITO/ZnO/TFB/ZnCdSe/ZnSeS QDs/ZnO/Al) using ZnO (**a**, +4 V) and SnO₂ (**b**, +8 V) with different thicknesses as the ETLs, respectively. The spectra of devices with 44 nm and without ZnO nanoparticles in Fig. 5a are enlarged by a factor of 2.7 and 12 for visibility, respectively. **c**, Schematic diagram of the electron injection process using ZnO as ETL. The energy of electrons in the Al electrode is decreased within the ZnO layer, so they are injected into the band edge of QDs, decreasing the electron leakage probability and increasing the lifetime of the device.

Hot electrons possess energy at least several hundreds of meV higher than that of the band edge electron, which facilitate electron leakage into HTL[10], possibly causing problems such as efficiency roll-off, instability issue of blue QD-LEDs, and inferior performance of InP-based QD. The current optimal QD-LED architecture could have been unintentionally developed for hot electron management but the underline mechanism hasn't been well understood. For example, without exception, all current high-performance QD-LEDs use $Zn_{1-x}Mg_xO$ or ZnO nanoparticles as the ETL[36-39]. However, the reason why only $Zn_{1-x}Mg_xO$ and ZnO nanoparticles allow efficient electron injection while the other ETLs, such as $SnO_2$ and $TiO_2$ do not, remains unclear. In the following content, we seek to provide a reasonable explanation.

We first investigated a type of high-performance CdSe-based QD-LED with a high external quantum efficiency (EQE) of 19.4% (Supplementary Fig. 13 and Table 2), we started from the device with optimal ZnO thickness, and then gradually decreased it until they were absent. We found that the hot electron population (with +4 V bias) increases and EQE decreases monotonously as we decrease the ZnO thickness (Fig. 5 and Supplementary Fig. 14). Especially, in the device without ZnO, which shows a very low EQE of 0.08%, electrons mainly occupy the higher energy instead of the band edge states (Supplementary Fig. 15). These results indicate one beneficial role of ZnO is to cool down the hot electrons before injecting them into the QDs, possibly through their abundant surface dangling bonds.

We also measured QD-LEDs using $SnO_2$ as the ETL, which show much lower EQEs and shorter lifetimes than those using ZnO ETLs (Supplementary Fig. 16 and Table 3)[39,40]. The near-zero band edge bleaching signal at +4 V indicates the electron injection is way less efficient than that of ZnO (Supplementary Fig. 17). As we increase the bias to +8 V, electrons are mainly injected into the higher energy states, especially in devices using thinner $SnO_2$ layers, resulting in easier electron leakage into HTL. These results also serve as additional evidence to support the presence of a strong phonon bottleneck, as the band edge bleaching signal is so small (0.4 mOD) when there are noticeable higher energy state electrons. These data also shed new light on why $SnO_2$ is less efficient than ZnO while serving as ETL.

In conclusion, we observed abundant hot electrons in both CdSe-and InP-based QD-LEDs using our latest developed EETA spectroscopy. The hot electron lifetime reaches 371 ns, in striking contrast to sub-picosecond measured by PETA. These findings shed new light on the operation mechanism of QD-LEDs, especially the mysterious role of ZnO-based nanoparticles, reminding us of the importance of hot electron management in future device engineering. These results also encourage further attempts on hot carrier solar cells – the hot electrons can be sufficiently long-lived if they are separated from holes.